 \documentclass{emulateapj}

\shorttitle{Extreme star-formation-rate densities at $z=3.4$ revealed by ALMA $20 \, {\rm mas}$ resolution imaging}
\shortauthors{Oteo et al.}

\usepackage{natbib}

\begin{document}


\title{ALMACAL II: Extreme star-formation-rate densities in a pair of dusty starbursts at $z = 3.442$ revealed by ALMA 20-milliarcsec resolution imaging}


\author{I.~Oteo\altaffilmark{1,2}, M.\,A.~Zwaan\altaffilmark{2},  R.\,J.~Ivison\altaffilmark{2,1}, I.~Smail\altaffilmark{3,4}, and A.\,D.~Biggs\altaffilmark{2}}

\affil{$^1$Institute for Astronomy, University of Edinburgh, Royal Observatory, Blackford Hill, Edinburgh EH9 3HJ UK}
\affil{$^2$European Southern Observatory, Karl-Schwarzschild-Str.\ 2, 85748 Garching, Germany}
\affil{$^3$Centre for Extragalactic Astronomy, Department of Physics, Durham University, South Road, Durham DH1 3LE UK}
\affil{$^4$Institute for Computational Cosmology, Department of Physics, Durham University, South Road, Durham DH1 3LE, UK}
\email{ivanoteogomez@gmail.com}


  
\begin{abstract}

We present ALMA ultra--high--spatial resolution ($\sim 20 \, {\rm mas}$ or $150 \, {\rm pc}$) observations of dust continuum at $920 \, {\rm \mu m}$ and $1.2 \, {\rm mm}$ in a pair of submm galaxies (SMGs) at $z = 3.442$, ALMACAL--1 (A--1: $S_{\rm 870 \mu m} = 6.5 \pm 0.2 \, {\rm mJy}$) and ALMACAL--2 (A--2: $S_{\rm 870 \mu m} = 4.4 \pm 0.2 \, {\rm mJy}$). These are the brightest and most luminous SMGs discovered so far in ALMACAL, a wide and deep (sub)mm survey, which is being carried out in ALMA calibrator fields and currently contains observations at sub-arcsec resolution down to an ${\rm r.m.s.}$ of $\sim 15 \,{\rm \mu Jy \, beam^{-1}}$ in more than 250 calibrators fields. The spectroscopic redshifts of A--1 and A--2 have been confirmed via serendipitous detection of up to nine emission lines, in three different ALMA bands. Our ultra-high-spatial resolution data reveal that about half of the star formation in each of these starbursts is dominated by a single compact clump (FWHM size of $\sim 350 \, {\rm pc}$). This structure is confirmed by independent datasets at $920 \, {\rm \mu m}$ and $1.2 \, {\rm mm}$. In A--1, two additional, fainter clumps are found. The star-formation rate (SFR) surface densities of all these clumps are extremely high, $\Sigma_{\rm SFR} \sim 1200$ to $\sim 3000 \, {M_\odot \, {\rm yr}^{-1} \, {\rm kpc}^{-2}}$, the highest found in high-redshift galaxies. There is a small probability that A--1 and A--2 are the lensed components of a background source gravitationally amplified by the blazar host. If this was the case, the effective radius of the dusty galaxy in the source plane would be $R_{\rm eff} \sim 40 \, {\rm pc}$, and the de-magnified SFR surface density would be $\Sigma_{\rm SFR} \sim 10000  \, {M_\odot \, {\rm yr}^{-1} \, {\rm kpc}^{-2}}$, comparable with the eastern nucleus of Arp 220. Despite being unable to rule out an AGN contribution, our results suggest that a significant percentage of the enormous far-IR luminosity in some dusty starbursts is concentrated in very small star-forming regions. The high $\Sigma_{\rm SFR}$ in our pair of SMGs could only be measured thanks to the ultra--high--resolution ALMA observations used in this work, demonstrating that long-baseline observations are essential to study and interpret the properties of dusty starbursts in the early Universe.

\end{abstract}


%

\section{Introduction}\label{intro}

Two decades ago, the first large format bolometer cameras on single-dish submm telescopes discovered a population of galaxies that were forming stars at tremendous rates, the so-called submm galaxies \citep[SMGs,][]{Smail1997ApJ...490L...5S, Barger1998Natur.394..248B, Hughes1998Natur.394..241H, Blain2002PhR...369..111B}. Later, it was reported that these starbursts were predominantly at high redshift, $z \sim 1 - 3$ \citep{Chapman2005ApJ...622..772C, Simpson2014ApJ...788..125S}. One of the main problems of these single-dish submm observations is their large beams, typically $>10''$. This complicates the multi-wavelength counterpart identification in the absence of higher-resolution (sub)mm follow-up and prevents us from studying the morphology of dust emission, needed to help interpret the properties of the ISM in dusty starbursts.

Interferometric observations at arcsec and sub-arcsec resolution revealed that most SMGs are major mergers, both from morphological and kinematics arguments \citep[e.g.][]{Tacconi2008ApJ...680..246T,Engel2010ApJ...724..233E}. Building on early indications from radio imaging \citep{Ivison2007MNRAS.380..199I}, ALMA revealed that single-dish submm sources are normally resolved into several distinct components \citep{Karim2013MNRAS.432....2K,Hodge2013ApJ...768...91H}, although it is not clear that all of them are at the same redshift and, therefore, physically associated. Based on limited ALMA data, \cite{Ikarashi2015ApJ...810..133I} reported that the dust in SMGs at $z > 3$ is confined to a relatively compact region, with a FWHM size of $\sim 0.2''$ or $\sim 1.5\,{\rm kpc}$. This average value is compatible with the size of SMGs at slightly lower redshifts reported in \cite{Simpson2015ApJ...799...81S}. Due to the still modest spatial resolution in those works, it was not possible to explore any sub-galactic structure within the SMGs. Using observations at higher spatial resolution ($\sim 0.1''$), \cite{Oteo2016arXiv160107549O} studied the morphology of two interacting starbursts at $z \sim 4.4$. The small beam size resolved the internal structure of the two sources, and revealed that the dust emission is smoothly distributed on $\sim {\rm kpc}$ scales, in contrast with the more irregular [CII] emission. 

Analysing strongly lensed sources offers an alternative to high-spatial-resolution observations \citep{Swinbank2010Natur.464..733S,Negrello2010Sci...330..800N,Bussmann2013ApJ...779...25B,Bussmann2015ApJ...812...43B}. Arguably, the best example is the ALMA study for SDP.81 \citep{Vlahakis2015ApJ...808L...4A}, a strongly lensed starburst at $z \sim 3$ \citep{Negrello2014MNRAS.440.1999N,Dye2014MNRAS.440.2013D,Frayer2011ApJ...726L..22F} selected from the {\it Herschel}-ATLAS \citep{Eales2010PASP..122..499E}. \cite{Dye2015MNRAS.452.2258D} modelled the lensed dust and CO emission of SDP.81 \citep[see also][]{Rybak2015MNRAS.451L..40R,Rybak2015MNRAS.453L..26R} and the dynamical analysis presented in \cite{Swinbank2015ApJ...806L..17S} revealed that SDP.81 comprises at least five star-forming clumps, which are rotating with a disk-like velocity field. However, with lensed galaxies the results (specially those lensed by galaxy-scale potential wells) must rely on accurate lens modeling ensuring that all the recovered source-plane emission is real and not an artifact of the modeling itself. Furthermore, and importantly, even relatively bright intrinsic emission can lie below the detection threshold if the geometry is not favourable, giving a misleading picture.

Thanks to the unique sensitivity and long-baseline capabilities of ALMA, ultra-high-spatial resolution observations can be carried out, for the first time, in unlensed FIR-bright sources. In this work we present ultra-high-spatial resolution observations ($\sim 20 \, {\rm mas}$) in a pair of submm galaxies (SMGs) at $z = 3.442$ selected from {\sc ALMACAL} \citep{Oteo2016ApJ...822...36O}. The main difference between our and previous work \citep{Simpson2015ApJ...799...81S,Ikarashi2015ApJ...810..133I} is the use of a significant number of very long baselines, providing $\sim 10\times$ better spatial resolution. Furthermore, our in-field calibrator and subsequent self-calibration ensures near-perfect phase stability on the longest baselines. Additionally, we have two independent datasets in ALMA band 6 (B6) and band 7 (B7), which prove the reliability of the structure we see. 

The paper is structured as follows: \S \ref{data_set_ALMACAL} presents the data set used in this work. \S \ref{section_redshift_confirmation_J1058} presents the redshift confirmation of our two sources and their FIR SED. In \S \ref{section_morphology_pc_scales} we discuss the morphology of the dust emission in our sources at $0.02''$ or $\sim 150\,{\rm pc}$ resolution. Finally, \S \ref{concluuuuu} summarizes the main conclusions of the paper. A \cite{Salpeter1955} IMF is assumed to derive star-formation rates (SFRs). assume a flat universe with $(\Omega_m, \Omega_\Lambda, h_0)=(0.3, 0.7, 0.7)$. For this cosmology, the angular scale is $\sim 7.3\,{\rm kpc}$ per arcsec at $z = 3.442$, the redshift of the sources under study.

\begin{figure}
\centering
\includegraphics[width=0.45\textwidth]{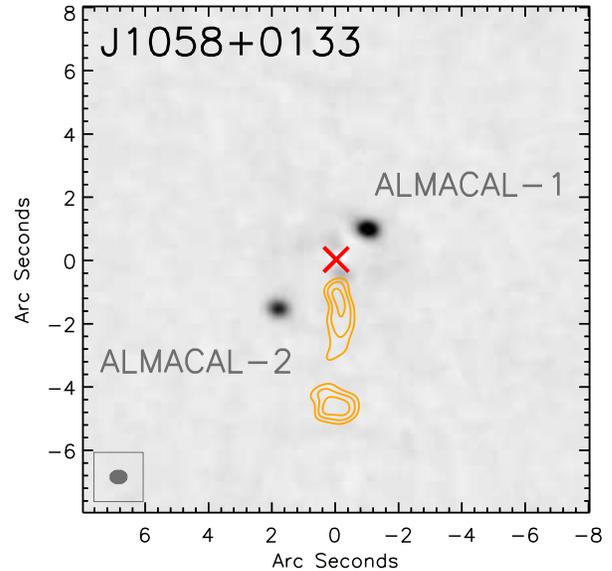}
\caption{Continuum map ($870 \, {\rm \mu m}$) of the two dusty starbursts at $z = 3.442$ (ALMACAL--1 and ALMACAL--2) discovered around the calibrator J1058+0133 at $z = 0.88$. The coordinates of the two sources can be found in Table \ref{table_properties_J1058_SMGs}. The calibrator has been subtracted from the data in the $uv$ plane by using a point--source model and is located at the position marked by the red cross. Orange contours represent the jet emanating from J1058+0133, revealed by $3 \, {\rm mm}$ imaging. The image is $16''$ on each side, and the beam of the $870 \, {\rm \mu m}$ continuum observations is shown on the bottom left.
              }
\vspace{5mm}
\label{fig_map_calibrator}
\end{figure}

\begin{figure*}
\centering
\includegraphics[width=0.35\textwidth]{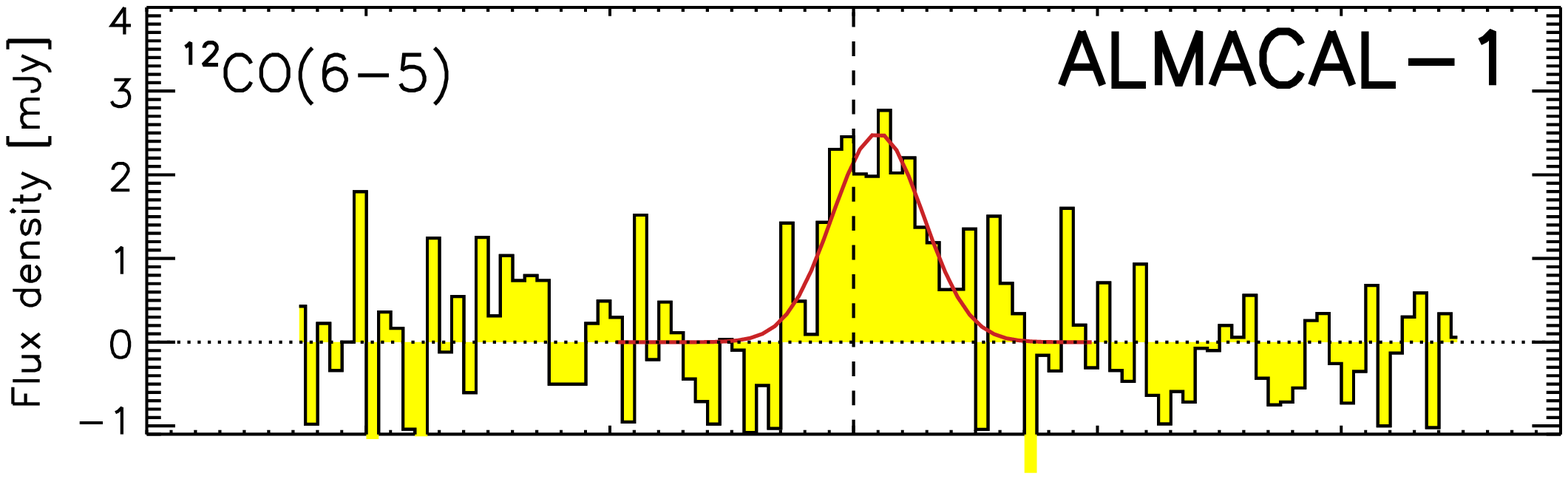}
\hspace{-11mm}
\includegraphics[width=0.35\textwidth]{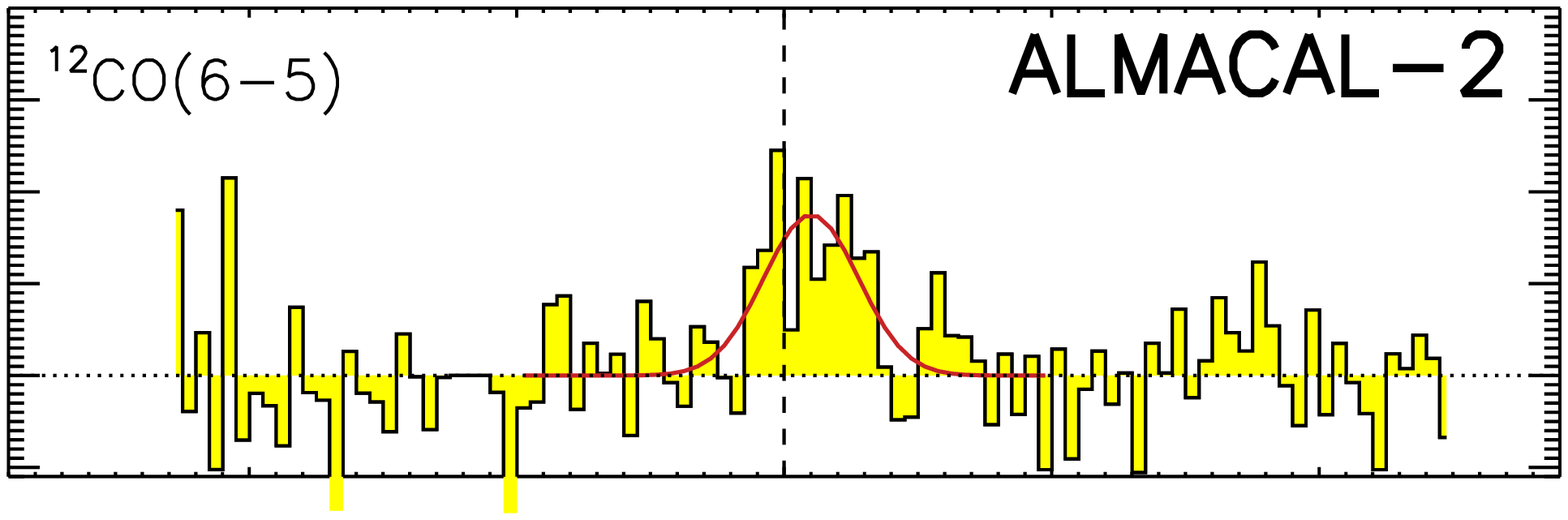}\\

\vspace{-4mm}
\includegraphics[width=0.35\textwidth]{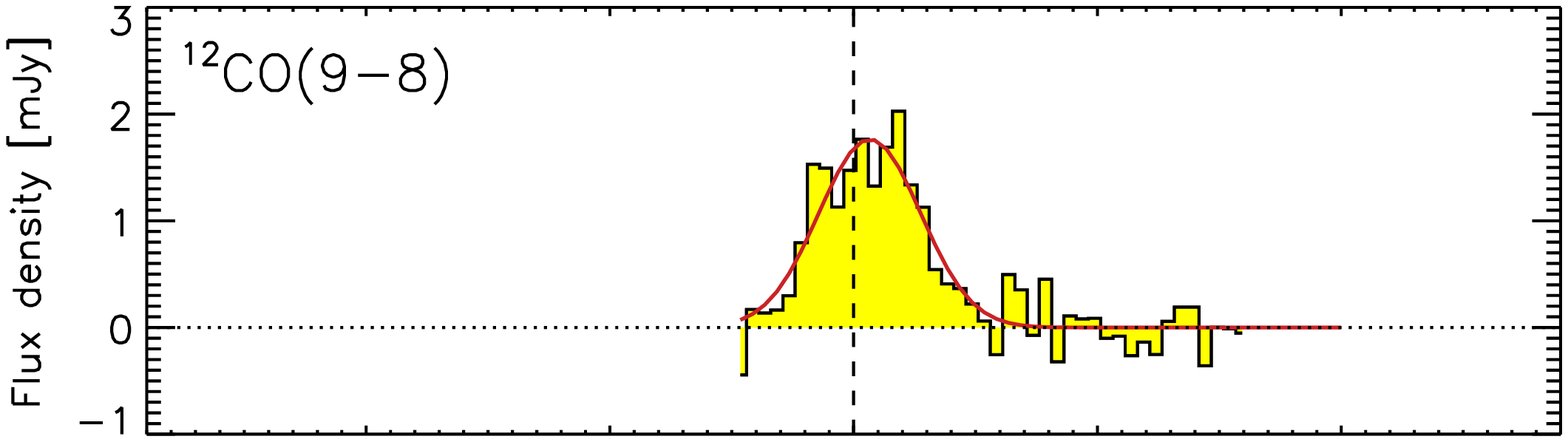}
\hspace{-11mm}
\includegraphics[width=0.35\textwidth]{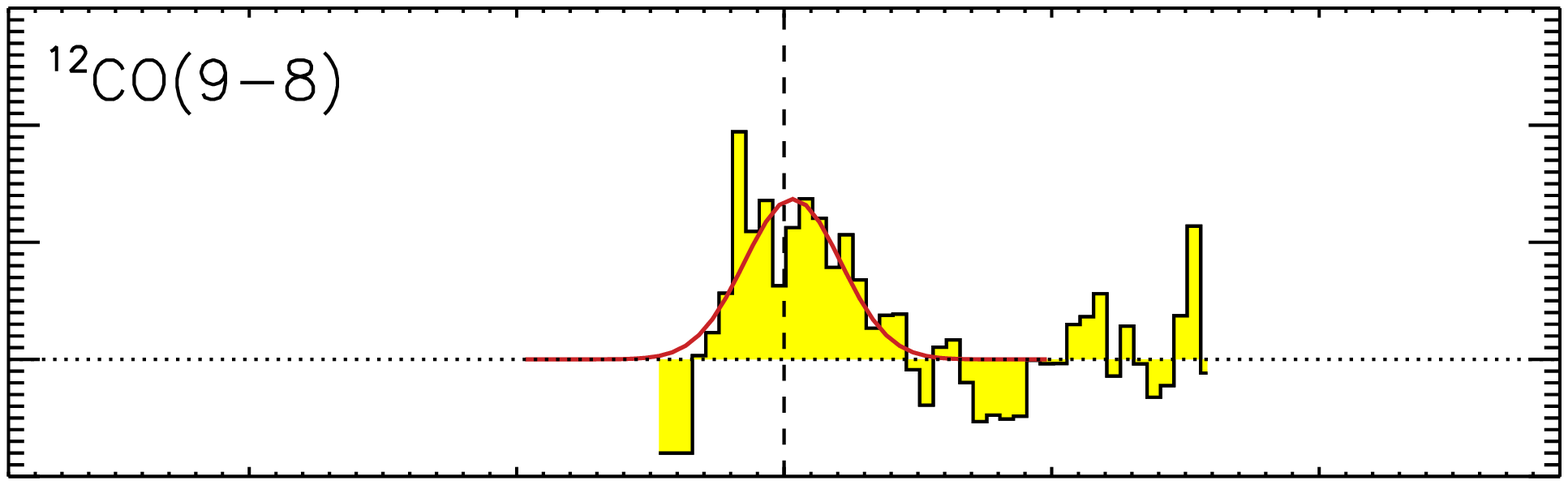}\\

\vspace{-4mm}
\includegraphics[width=0.35\textwidth]{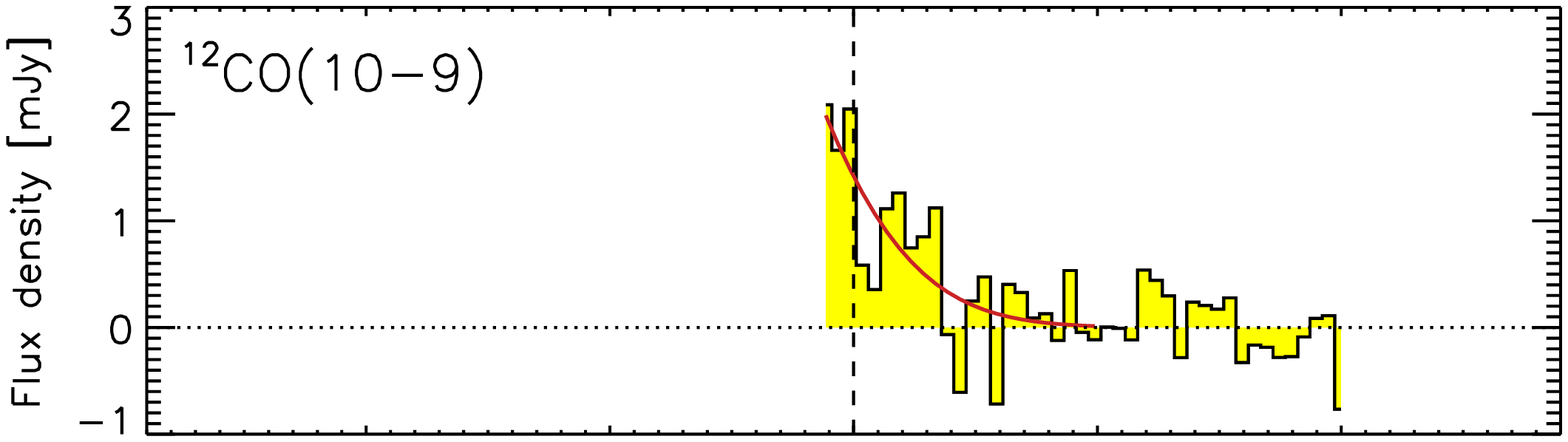}
\hspace{-11mm}
\includegraphics[width=0.35\textwidth]{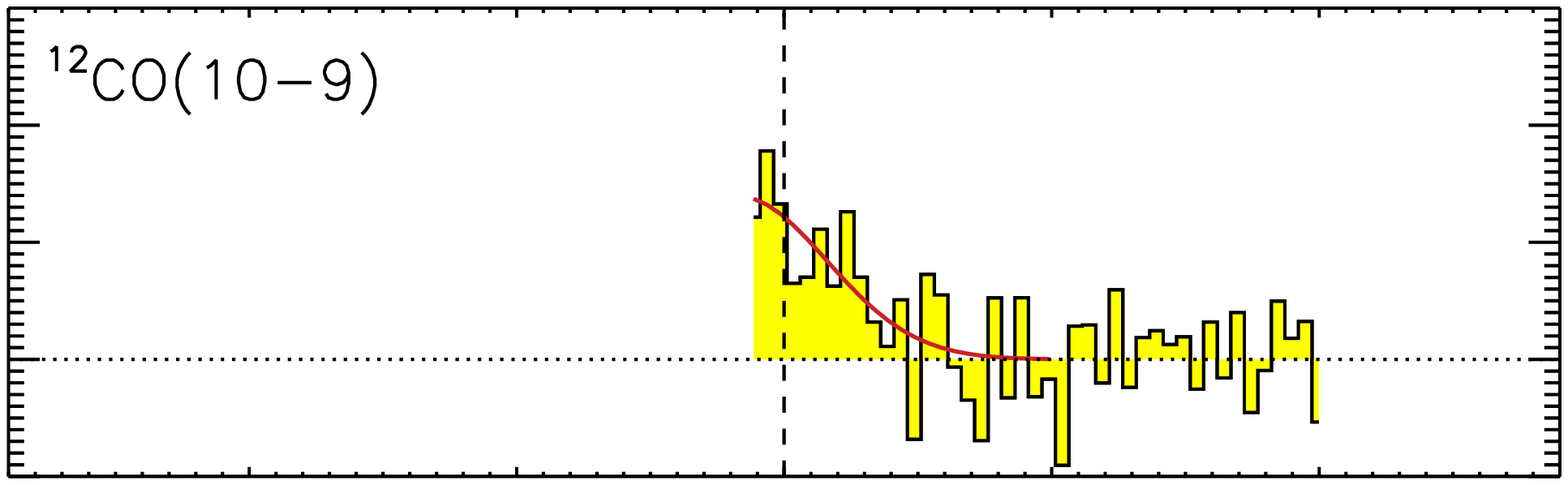}\\

\vspace{-4mm}
\includegraphics[width=0.35\textwidth]{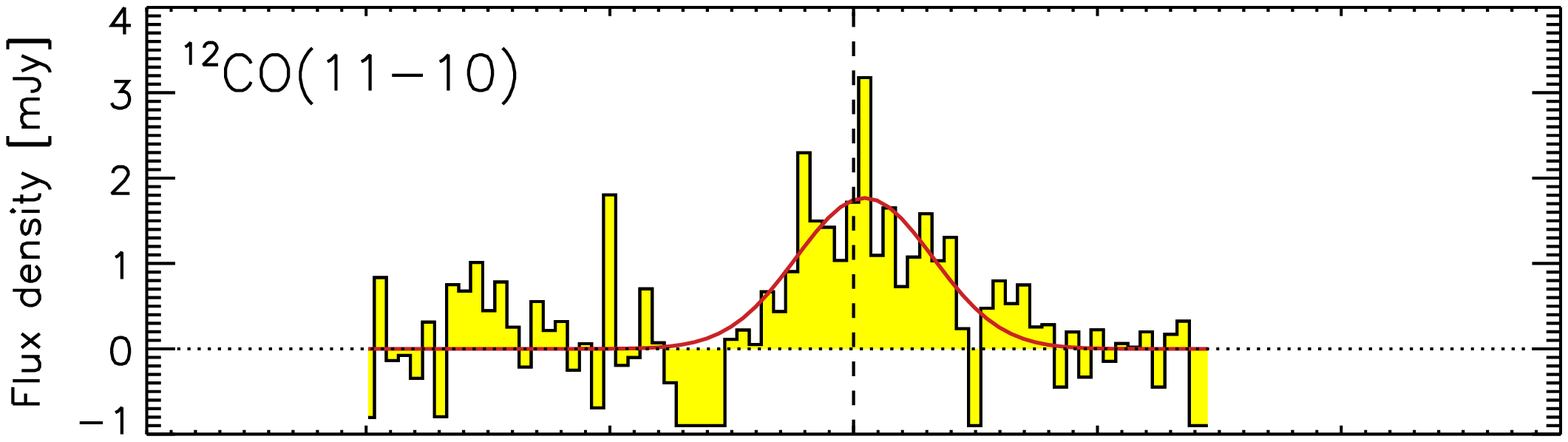}
\hspace{-11mm}
\includegraphics[width=0.35\textwidth]{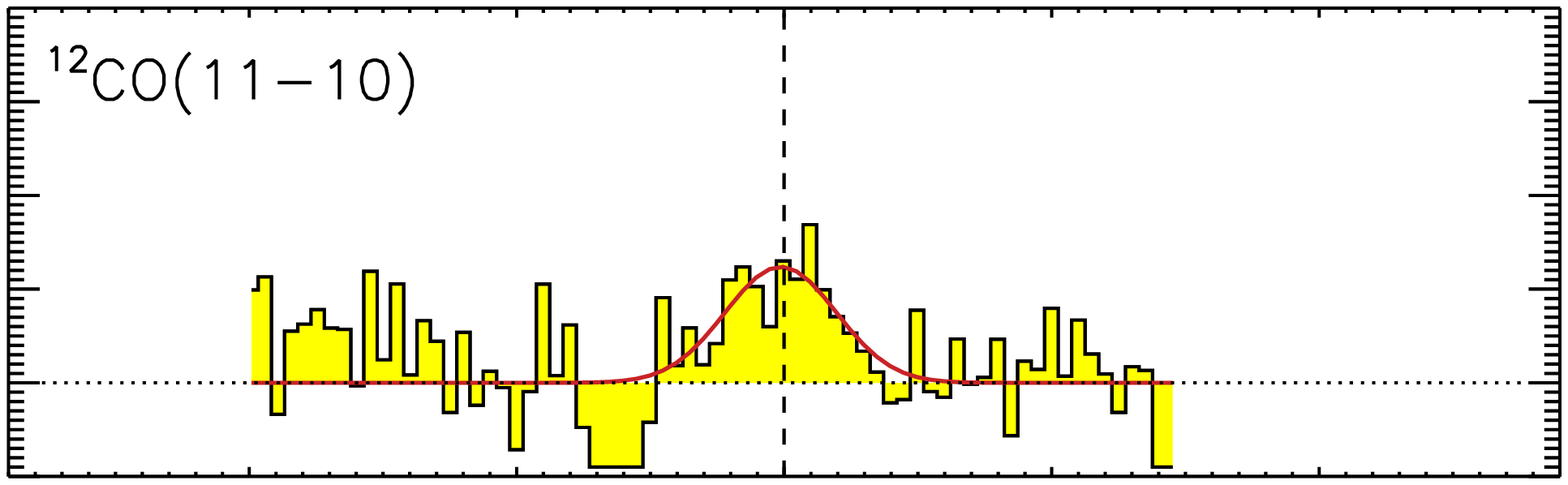}\\

\vspace{-4mm}
\includegraphics[width=0.35\textwidth]{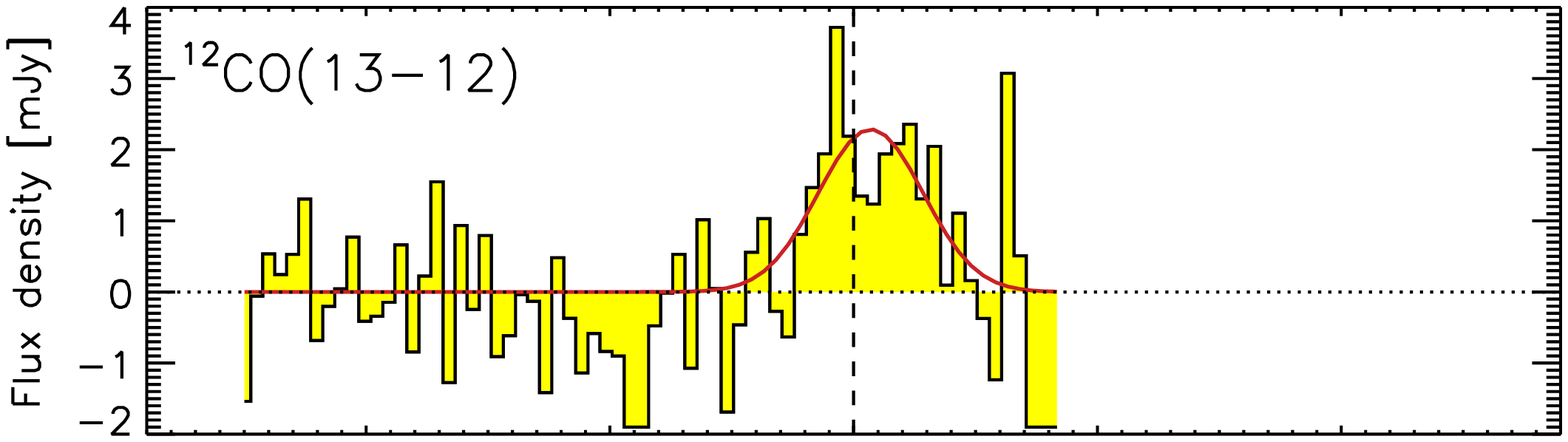}
\hspace{-11mm}
\includegraphics[width=0.35\textwidth]{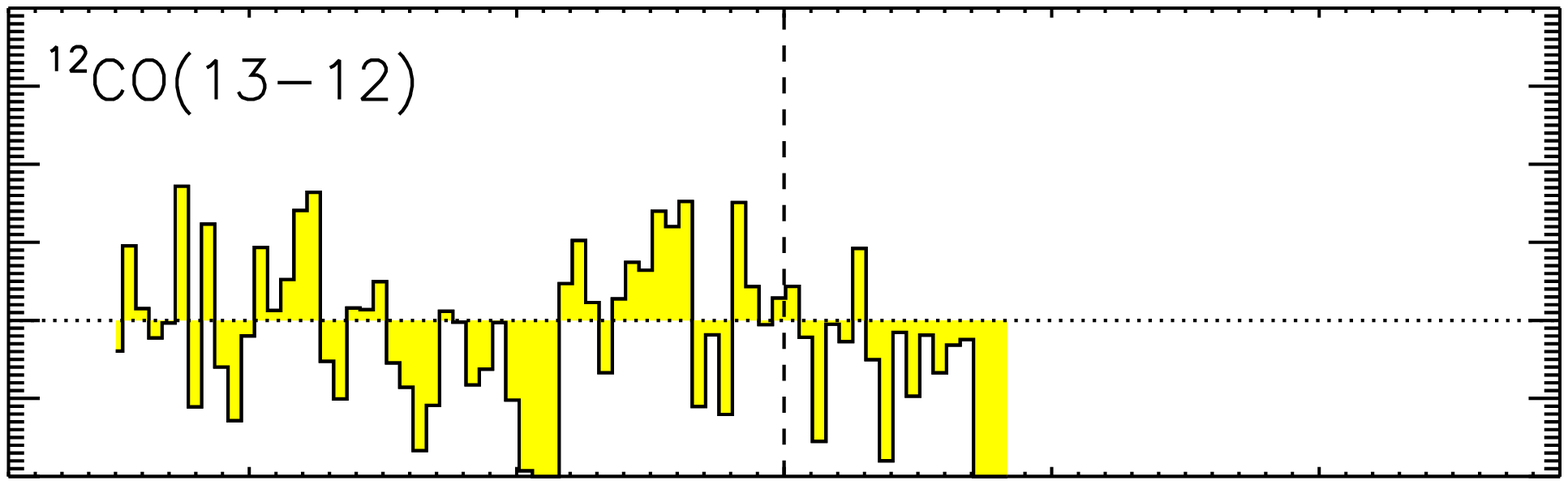}\\

\vspace{-4mm}
\includegraphics[width=0.35\textwidth]{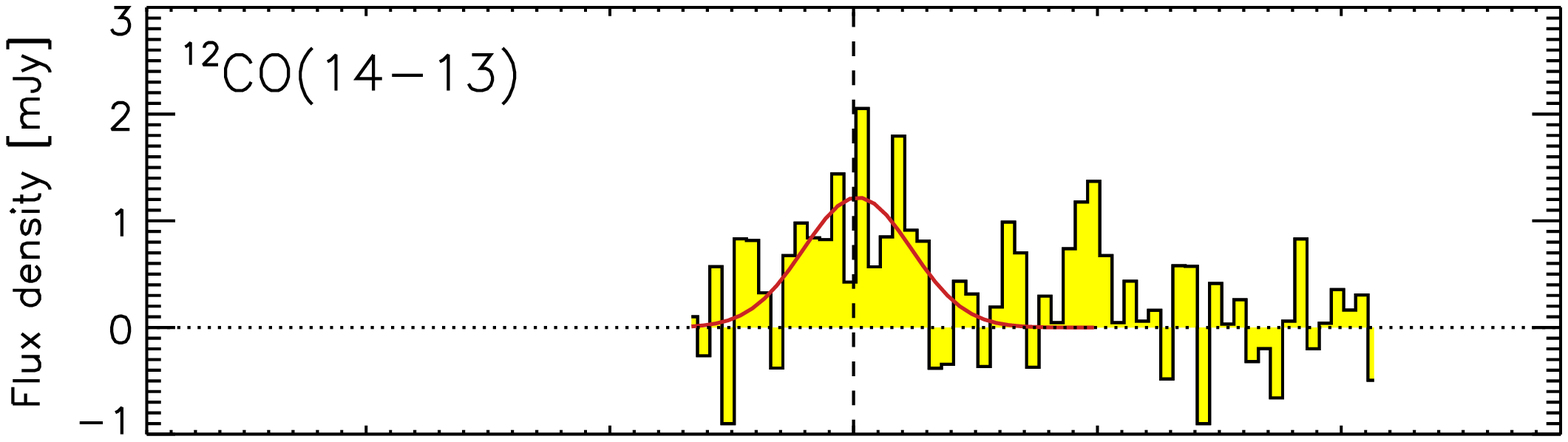}
\hspace{-11mm}
\includegraphics[width=0.35\textwidth]{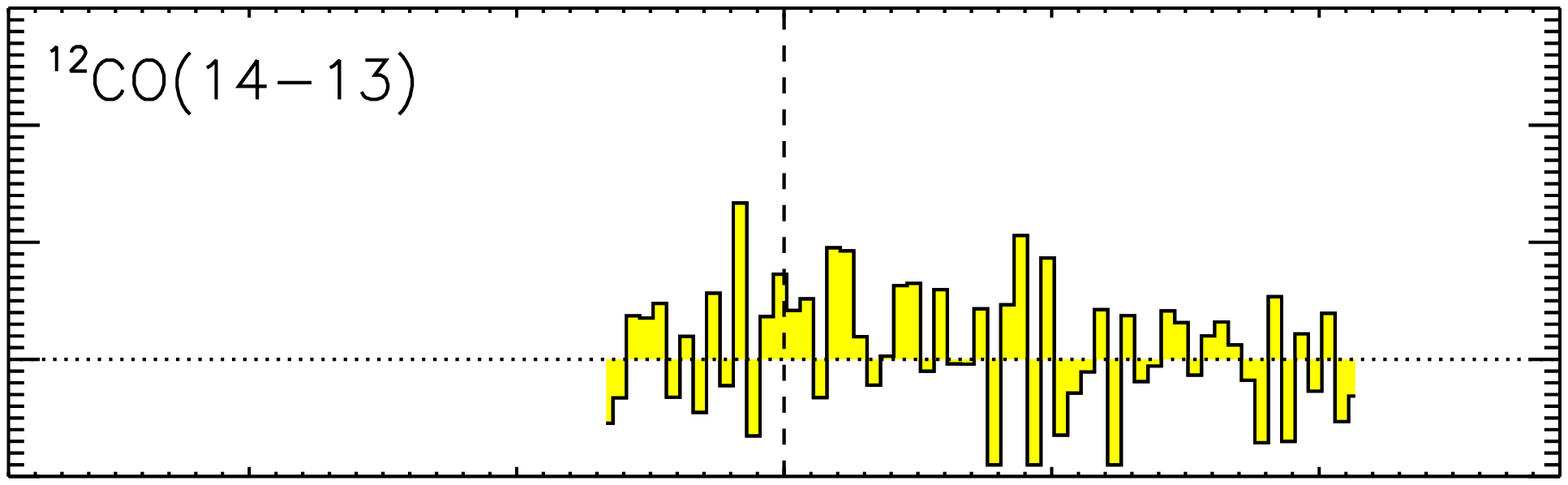}\\

\vspace{-4mm}
\includegraphics[width=0.35\textwidth]{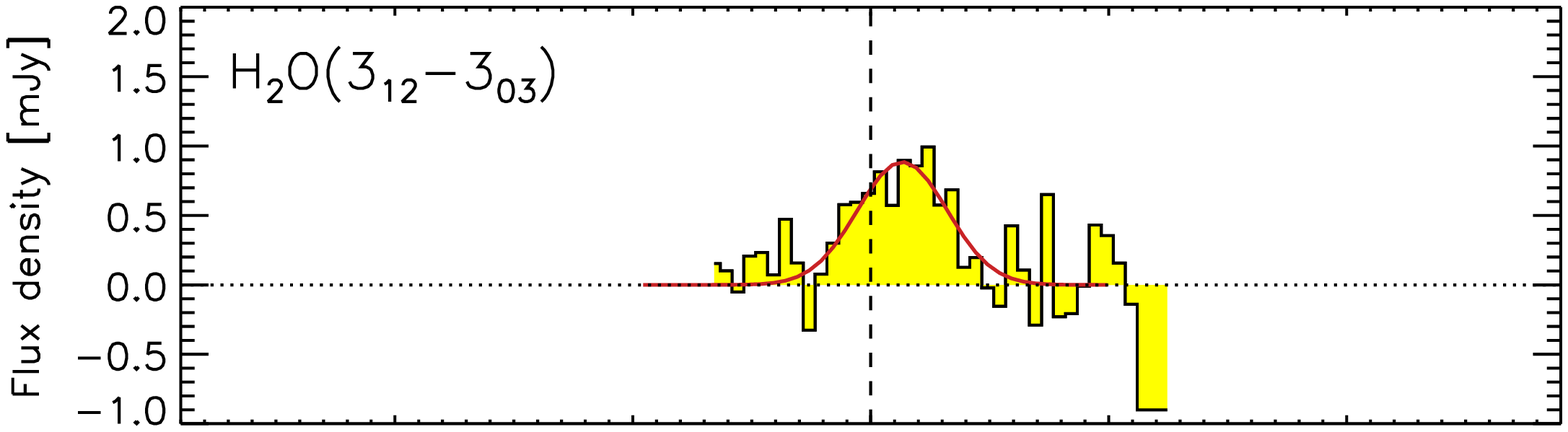}
\hspace{-11mm}
\includegraphics[width=0.35\textwidth]{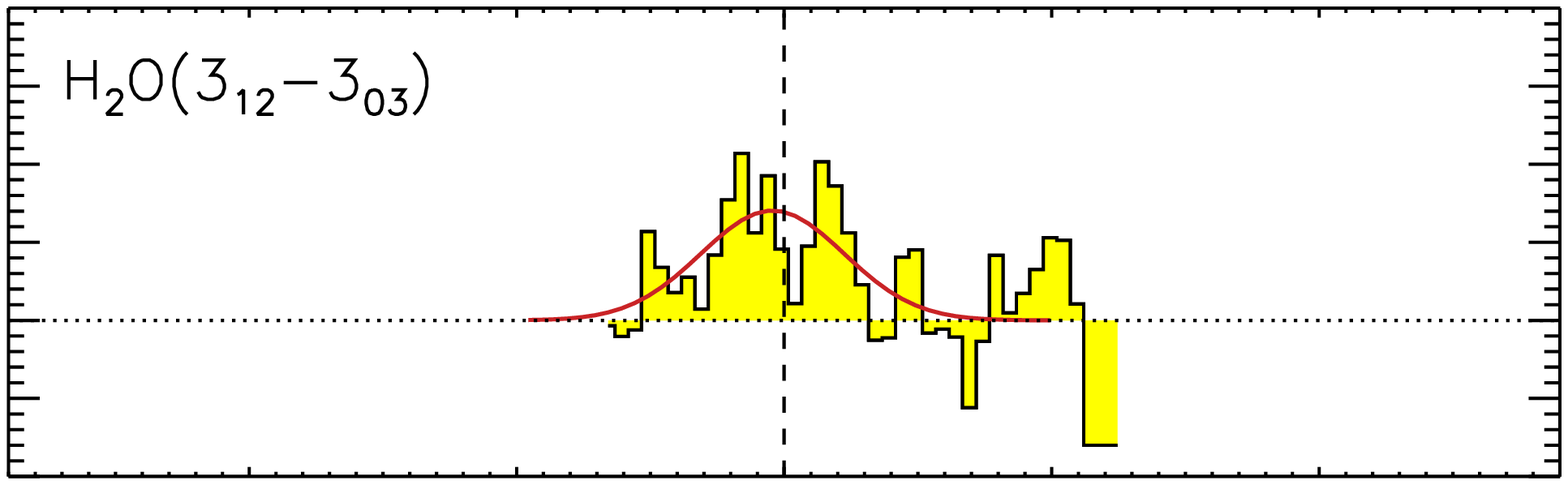}\\

\vspace{-4mm}
\includegraphics[width=0.35\textwidth]{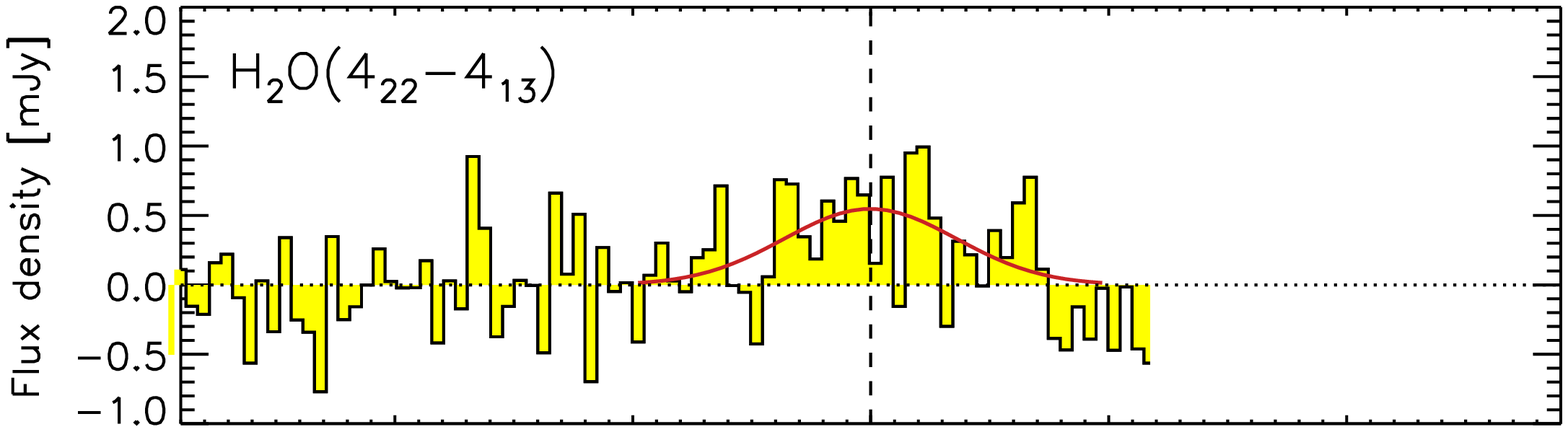}
\hspace{-11mm}
\includegraphics[width=0.35\textwidth]{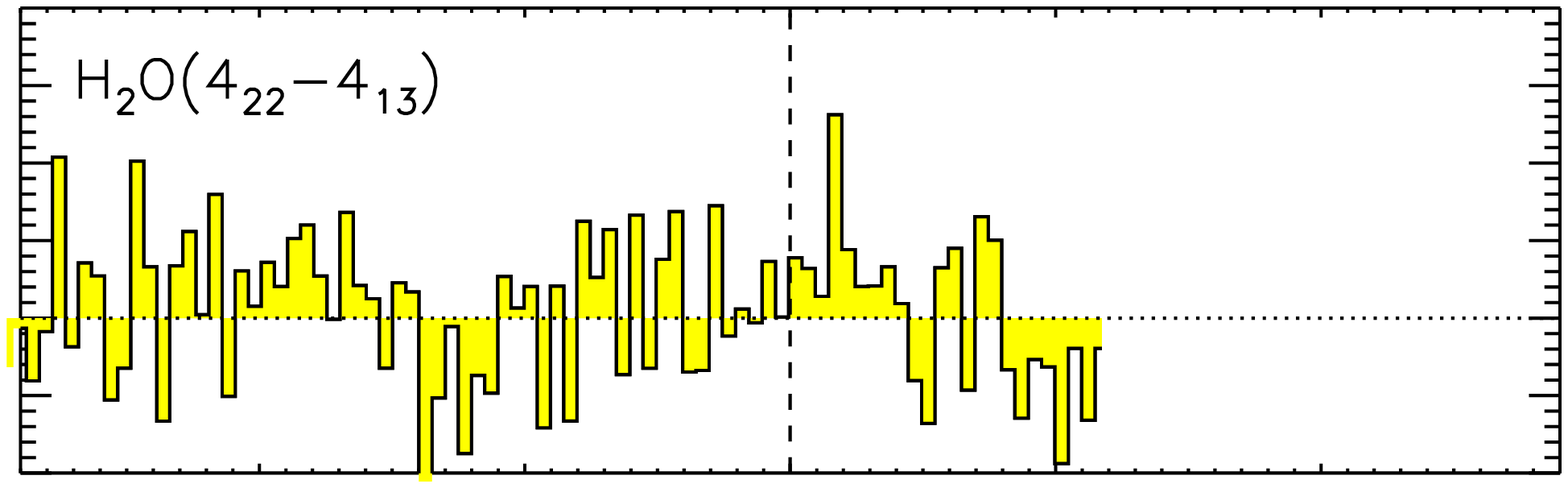}\\

\vspace{-4mm}
\includegraphics[width=0.35\textwidth]{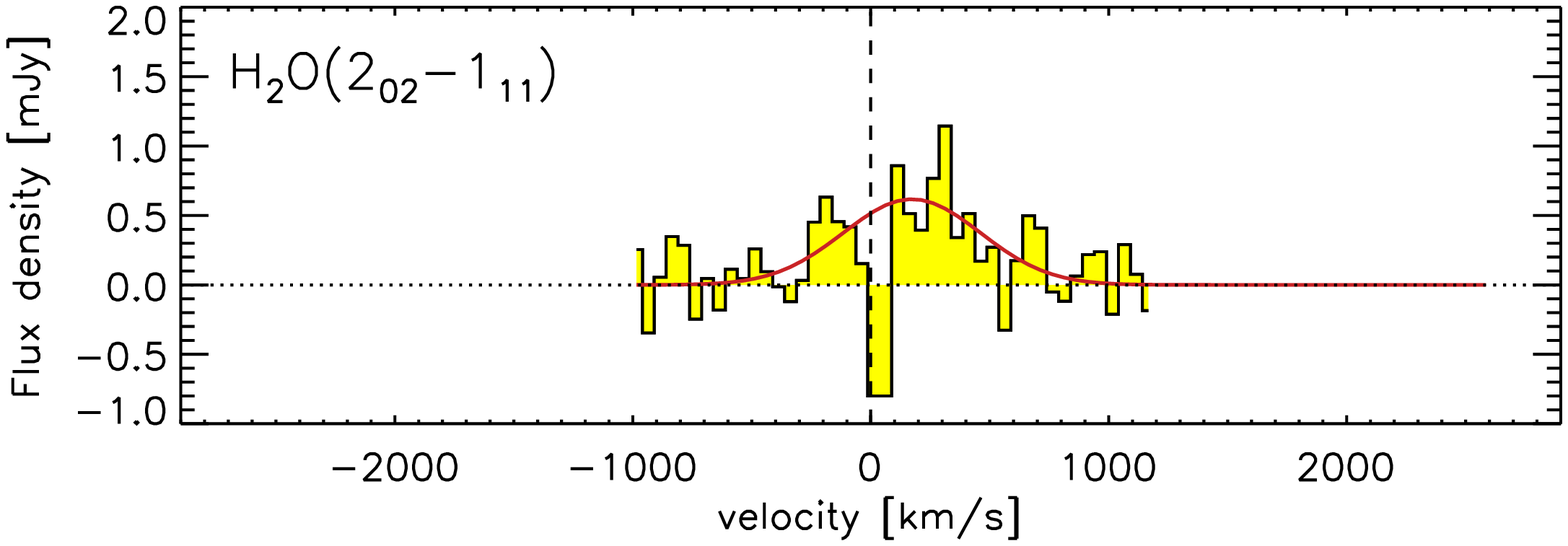}
\hspace{-11mm}
\includegraphics[width=0.35\textwidth]{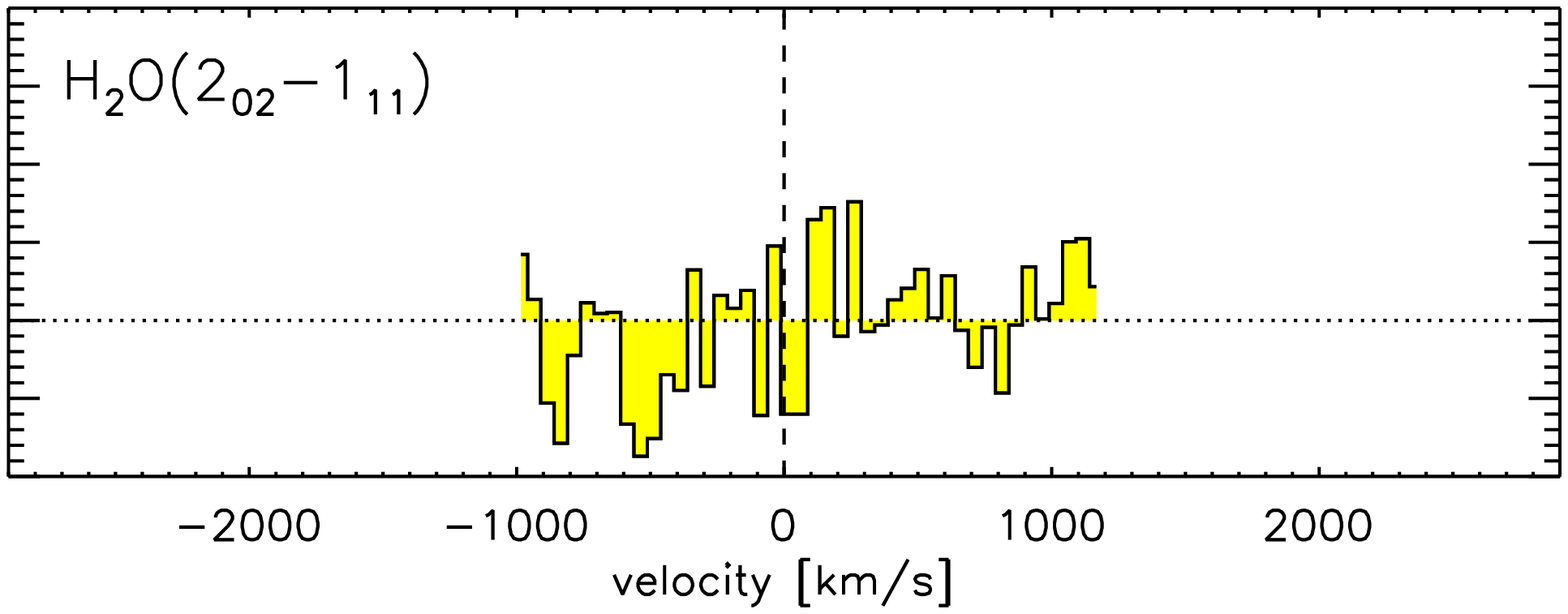}\\

\caption{Continuum-substracted spectra showing the coverage of emission lines in our two SMGs, ALMACAL--1 ({\it left}) and ALMACAL--2 ({\it right}). Up to nine emission lines are detected in each source, unambiguously confirming that their redshift is $z = 3.442$. The detected emission lines (except $^{12}$CO(10-9) which is only half covered) are fitted with Gaussian profiles (plotted as the red curves) in order to calculate their observed fluxes. The absence of a Gaussian fit in a given panel means that the corresponding line has not been detected. The vertical dashed lines indicate $v = 0 \, {\rm km \, s^{-1}}$ for a redshift $z = 3.442$. It should be pointed out that the redshift confirmation has been obtained from high-$J$ CO and ${\rm H_2O}$ lines, not usual for FIR-bright sources, where redshift confirmation is normally achieved using spectral scans in the 3mm band \citep{Weiss2009ApJ...705L..45W_CO,Weiss2013ApJ...767...88W,Asboth2016arXiv160102665A,Oteo2016arXiv160107549O,Strandet2016arXiv160305094S}.
              }
\label{detected_lines_SMGs}
\end{figure*}


%
%

\begin{figure}
\centering
\includegraphics[width=0.48\textwidth]{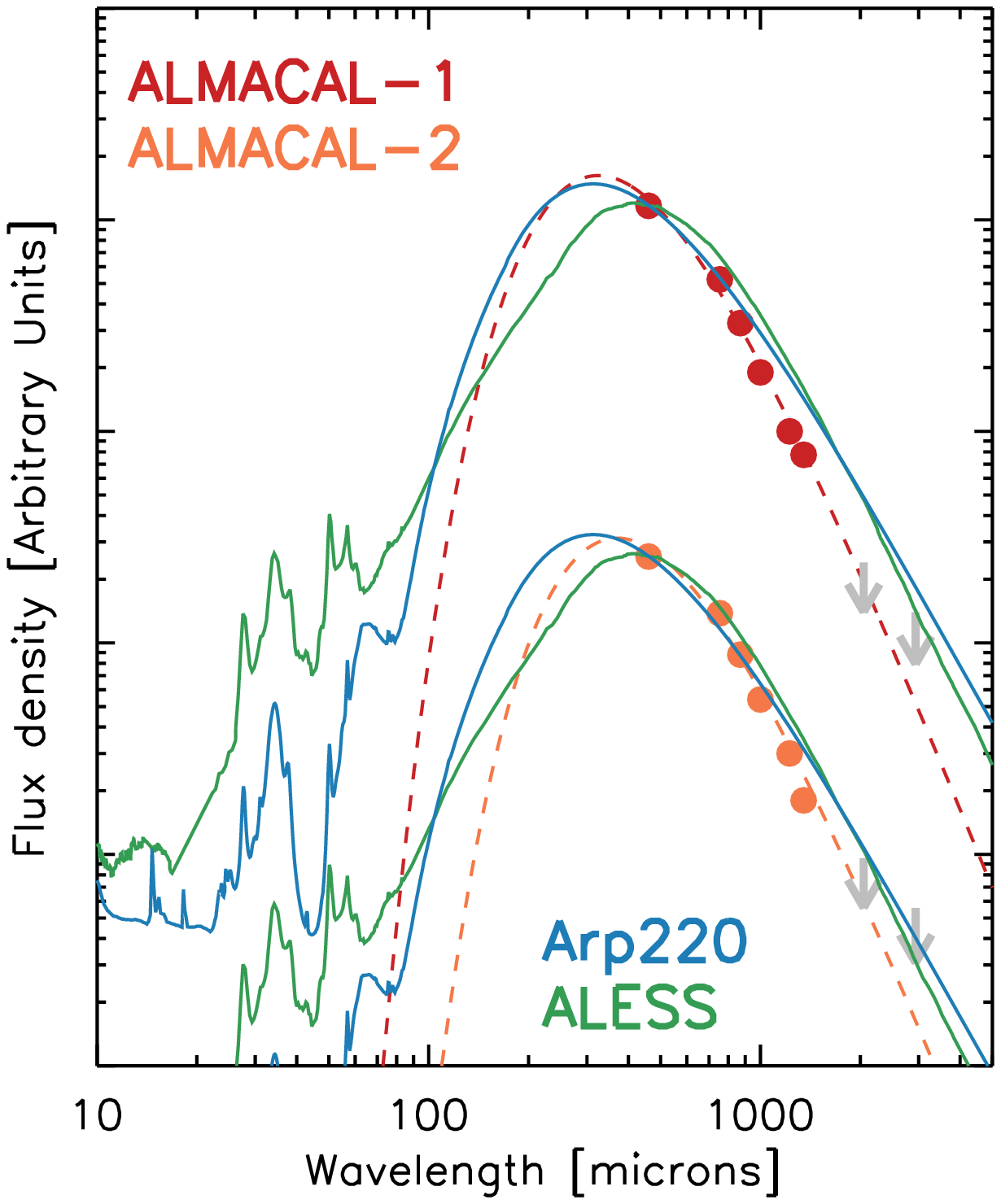}
\caption{FIR SED of A--1 (red) and A--2 (orange). All photometric points come from the multi-band observations in ALMA bands 6, 7, 8 and 9. Since there are available data on each side of B6 and B7 we have split the data for those bands in two sub-bands. With this, we have six photometric points for each source (and two $5 \sigma$ upper limits in ALMA bands 4 and 3 indicated by the grey arrows). It should be noted that the error bars on the photometric points are smaller than the size of the filled dots. The FIR SEDs have been fitted assuming optically thin models with dust emissivity $\beta = 2.0$ (dashed curves) to derive their dust temperature, and total IR luminosities (see Table \ref{table_properties_J1058_SMGs}). For a reference, we have included the templates associated to the average FIR SED of ALESS SMGs \citep{Swinbank2014MNRAS.438.1267S} and Arp 220, redshifted to $z = 3.442$ and re-scaled using the observed $460 \, {\rm \mu m}$ flux density of each source.
              }
\vspace{5mm}
\label{figure_TDLIR_components}
\end{figure}

\section{Data set: ALMACAL}\label{data_set_ALMACAL}

Using ALMA calibration data we are carrying out a wide and deep (sub)mm survey, ALMACAL. The strategy of the survey and the details of data calibration and source extraction can be found in \cite{Oteo2016ApJ...822...36O}. Briefly, our survey takes advantages of the huge amount of ALMA calibration data, which are routinely acquired during the execution of ALMA science projects. Combining compatible data for different calibrators it is possible to cover areas large enough and r.m.s. levels low enough to enable the detection of faint SMGs. 

At the present stage of the survey we are reaching noise levels down to ${\rm \sim 15 \mu Jy \, beam^{-1}}$ at sub-arc resolution in more than 250 calibrator fields, representing an area of more than 20 sq arcmin (Oteo et al. in prep). We focus this paper on the two SMGs found around the calibrator J1058+0133: ALMACAL--1 (A--1) and ALMACAL--2 (A--2), see Figure \ref{fig_map_calibrator}), which are the brightest SMGs found so far in ALMACAL.

The ALMACAL data used in this work can be classified in two different groups according to the spatial resolution they provide. On one hand we use mid--spatial--resolution data (beam sizes typically larger than $0.3''$) which are part of the automated ALMACAL data extraction and calibration and are used to measure the unresolved dust continuum emission in our two SMGs and to look for emission lines to confirm their redshift (see \S \ref{section_redshift_confirmation_J1058}). Due to their compact nature, A--1 and A--2 remain unresolved in the mid-resolution data. 

In addition, and with the aim of studying the morphology of the dust emission in our two SMGs (\S \ref{section_morphology_pc_scales}), we also use ultra--high--spatial--resolution data especially extracted from the ALMA archive for the analysis presented in this work. Since no bright emission lines are covered by the spectral setup of the ultra--high--spatial--resolution observations, we focus on the continuum dust emission. There are ultra--high--spatial--resolution observations in B3, B6 and B7. No continuum emission is detected in B3 due to the lack of depth. We thus focus our analysis on B6 and B7. 

The extraction and calibration of the ultra-high resolution data is done in exactly the same way as for the mid-resolution data, including self-calibration to improve image quality. Using Briggs weighting with a \verb+robust+ parameter equal to 0.5 we obtain a beam size of $25 \, {\rm mas} \times 18 \, {\rm mas}$ at $\sim 920 \, {\rm \mu m}$ (B7), meaning a spatial resolution of $180 \, {\rm pc} \times 130 \, {\rm pc}$ at the redshift of the two sources. The continuum sensitivity is $\sigma_{\rm B7} = 30 \, {\rm \mu Jy} \, {\rm beam}^{-1}$. The spatial resolution of the B6 observations is $29 \, {\rm mas} \times 25 \, {\rm mas}$, with a r.m.s. level of $\sigma_{\rm B6} = 40 \, {\rm \mu Jy} \, {\rm beam}^{-1}$. The spatial resolution provided by these observations is about 10$\times$ times better than those reported so far in any previous unlensed, high-redshift starburst, and close to the typical size of giant molecular clouds ($\sim 50 \, {\rm pc}$).


\section{A pair of SMGs at $z = 3.442$}\label{section_redshift_confirmation_J1058}

As pointed out in \cite{Oteo2016ApJ...822...36O}, one of the key advantages of using ALMA calibrators to study the submm galaxy population is that they are typically observed in multiple ALMA bands. This allows us to: (1) have a good sampling of the FIR SED of the detected galaxies; (2) find redshifts by carrying out blind searches of (sub)mm emission lines (including CO, ${\rm H_2O}$, [CI], [CII], etc). The two SMGs found around calibrator J1058+0133 perfectly exemplify these two points. They were initially discovered in B6 and B7 as two bright sources near J1058+0133, a bright blazar at $z \sim 0.88$ used routinely as an ALMA calibrator \citep{Oteo2016ApJ...822...36O}. We thought initially that they were part of a jet emanating from the calibrator. However, ALMA B3 data revealed that J1058+0133 does have a strong jet, but not in the direction from the calibrator to either of the two SMGs (see Figure \ref{fig_map_calibrator}). The flux density ratio between $870 \, {\rm \mu m}$ and $1.2 \, {\rm mm}$ of each source (in addition to the lack of continuum detection in B3 and B4) was compatible with them being high-redshift SMGs, although it could still be compatible with the two sources being companions of the bright calibrator, located at its redshift. 

We then searched for emission lines from the two SMGs assuming that they were at the same redshift of the calibrator. Nothing was found. However, a blind search for emission lines in their (sub)mm spectrum revealed two emission lines in each component, unambiguously confirming a redshift of $z \sim 3.442$. Further data for this calibrator was then extracted from the ALMA archive, and up to nine potential emission lines were identified in each component, as shown in Figure \ref{detected_lines_SMGs}. We have detected $^{12}$CO(14--13), $^{12}$CO(13--12), $^{12}$CO(11--10), $^{12}$CO(10--9), $^{12}$CO(9--8), $^{12}$CO(6--5), H$_2$O(3$_{12}$--3$_{03})$ and weak H$_2$O(4$_{22}$--4$_{13})$ and H$_2$O(2$_{02}$--1$_{11})$ transitions in one or both sources. The median line width of the lines in A--1 and A--2 are $520$ and $417\,{\rm km \, s^{-1}}$, respectively, and there is evidence that our measured water lines are slightly wider than the CO transitions; this will be reassessed as more data for J1058+0133 become available. For a given source, the velocity shifts can be as large as $\sim 100 \, {\rm km \, s^{-1}}$. These are lower than those found in other bright starbursts such as SGP38326 \citep{Oteo2016arXiv160107549O} or HATLAS J084933 \citep{Ivison2013ApJ...772..137I}. 

The CO SLED of A--1 and A--2, including data for lower-$J$ transitions from other facilities where J1058+0133 has also been used as a calibrator (for example the JVLA), will be presented in a subsequent paper. However, it is important to point out here that A--1 is warmer (see \S \ref{section_FIR_SEDs} and Table \ref{table_properties_J1058_SMGs}), has relatively bright $^{12}$CO(13--12), $^{12}$CO(14--13) and ${\rm H_2O}(4_{22}-4_{13})$ lines, and its CO SLED seems to plateau at $J = 10-13$, suggesting an influential AGN in A--1 and less so in A--2, for which the upper limits in the high-$J$ CO lines suggest a less excited CO SLED. Despite the possible influence of an AGN on the molecular line properties, with the data in hand it is not possible to estimate the contribution of the possible AGN to the total IR luminosity of the source.

The redshift of A--1 and A--2 (see Table \ref{table_properties_J1058_SMGs}) clearly indicates that they are not related to the calibrator (at $z \sim 0.88$). It might be argued that the two SMGs are lensed components of the same background galaxy, as is suggested by the apparent symmetry of the two SMGs with respect to the calibrator. However, despite the high signal to noise of the multi-band continuum detections, there is no sign of an Einstein ring or extended emission connecting the two sources, sometimes seen in the lensed dust emission of high-redshift SMGs \citep[see for example][]{Vlahakis2015ApJ...808L...4A,Dye2015MNRAS.452.2258D,Bussmann2013ApJ...779...25B,Bussmann2015ApJ...812...43B}. Also, $^{12}$CO(13--12) and $^{12}$CO(14-13) are not detected in A--2 despite the low r.m.s. of the spectra would have allowed detections if the CO SLEDs of A--1 and A--2 were the same, as expected if they A--1 and A--2 were lensed components of the same background source. Furthermore, the FIR SED of the two sources are different (see \S \ref{section_FIR_SEDs}), and this is not compatible with them being lensed by the blazar host galaxy. Despite the arguments supporting the idea that A--1 and A--2 are not lensed, we will explore the consequences of possible lensing in the conclusions of this paper in \S \ref{lensing_section_discussion}.

A--1 and A--2 are separated by 28 kpc in projection, suggesting tidal interaction may have triggered star formation in both systems. If observed with a single-dish submm telescope, our two SMGs would have appeared in the image as a single unresolved blob. The separation between A--1 and A--2 is about 2$\times$ times the projected separation between the two interacting components of SGP38326 at $z = 4.425$ \citep{Oteo2016arXiv160107549O} and comparable to the separation between merging the HyLIRGs at $z \sim 2.4$ in \cite{Ivison2013ApJ...772..137I}. The $3.8''$ separation is compatible with the distance between the multiple components that SMGs are normally resolved into, as revealed by high-resolution radio or ALMA observations \citep{Ivison2007MNRAS.380..199I,Karim2013MNRAS.432....2K,Hodge2013ApJ...768...91H,Simpson2015ApJ...799...81S}.


\subsection{The far-IR SEDs}\label{section_FIR_SEDs}

In order to determine the dust temperature of A--1 and A--2 we have fitted their FIR SED (using all available photometry in B6, B7, B8 and B9) with optically thin models (see Figure \ref{figure_TDLIR_components}). Uniquely, we have performed FIR SED fits with sub-arcsec resolution photometry, unlike all previous work on high-redshift SMGs where the large beams of the single-dish observations prevents accurate deblending of the multiple components which SMGs are typically resolved into \citep{Karim2013MNRAS.432....2K,Hodge2013ApJ...768...91H,Simpson2015ApJ...799...81S}. It should be noted that we assume here that the total IR luminosity is due to star formation rather than AGN activity.

Since observations are available in almost all frequencies covering B6 and B7, we have split the observations in each band into two sub-bands corresponding to the two halves of each band. In this way, we have six photometric points in total (see values in Table \ref{table_properties_J1058_SMGs}), and a finer coverage of the FIR SED. Table \ref{table_properties_J1058_SMGs} quotes the dust temperature derived for A--1 and A--2 by assuming a dust emissivity of $\beta = 2.0$. It should be noted that lower $\beta$ values would give higher dust temperatures (for example, A--1 would have $T_{\rm D} = 48.0 \pm 1.4 \, {\rm K}$ for $\beta = 1.5$). However, the $\chi^2$ of the fits would not be significantly different and additional photometric information is required to distinguish between different values of $\beta$. The observed FIR SEDs suggests that A--1 is warmer than A--2.

In order to derive the total IR luminosity of each source (see Table \ref{table_properties_J1058_SMGs}) we have fitted their FIR SEDs using optically thin models (with a dust emissivity of $\beta = 2.0$), including a mid-IR power law with a slope of $\alpha = 2.25$. This provides a mid-IR SED similar to the one found for the average SMG population. The SFR of A--1 and A--2 has been then derived from the total IR luminosity assuming the classical \cite{Kennicutt1998} calibration and a Salpeter IMF (see Table \ref{table_properties_J1058_SMGs}). As expected from their brightness, the SFR of our two SMGs is very high, revealing extreme star formation and placing A--1 and A--2 among the most luminous starbursts at $z \sim 3-4$. The $T_{\rm D}$ and $L_{\rm IR}$ (see Table \ref{table_properties_J1058_SMGs}) of A--1 and A--2 are compatible to those found for the classical population of single-dish-submm-detected SMGs \citep{Swinbank2014MNRAS.438.1267S,Simpson2014ApJ...788..125S}, and they would have been selected individually as SMGs is they had been located in cosmological fields where FIR/(sub)mm surveys have been carried out. 

Figure \ref{figure_TDLIR_components} compares the FIR SED of A--1 and A--2 with the ones for the average population of SMGs in the ALESS survey \citep{Swinbank2014MNRAS.438.1267S} and Arp 220. These two templates have been shifted to $z = 3.442$ and scaled to the $460 \, {\rm \mu m}$ flux density of A--1 and A--2. It can be seen that the observed FIR SED of A--1 and A--2 are fainter at mm wavelengths for the same FIR flux density than Arp 220 and ALESS, suggesting that A--1 and A--2 are warmer than the average SMG (but still comparable with the spread of the $T_{\rm dust} - L_{\rm IR}$ relation). 
 
\begin{table}[!t]
\caption{\label{table_properties_J1058_SMGs}Observed properties of the two SMGs detected around calibrator J1058+0133}
\centering
\begin{tabular}{ccc}
\hline
 & A--1 & A--2 \\
\hline
RA & 10:58:29.7 & 10:58:29.5 \\
Dec & +1:33:57.2 & +1:33:59.7 \\
$z$ & $3.4433 \pm 0.0005$ & $3.4431 \pm 0.0005$\\
$S_{\rm 460\mu m}$ [mJy]		&	$23.3 \pm 1.3$ & $12.8 \pm 0.8$ \\
$S_{\rm 750\mu m}$ [mJy]		&	$10.5 \pm 0.6$ & $ 6.9 \pm 0.3$ \\
$S_{\rm 870\mu m}$ [mJy]		&	$6.5 \pm 0.2$ & $ 4.4 \pm 0.2$ \\
$S_{\rm 1000\mu m}$ [mJy]	&	$3.8 \pm 0.2$ & $ 2.7 \pm 0.2$ \\
$S_{\rm 1225\mu m}$ [mJy]	&	$2.0 \pm 0.1$ & $ 1.5 \pm 0.1$ \\
$S_{\rm 1350\mu m}$ [mJy]	&	$1.8 \pm 0.2$ & $ 0.9 \pm 0.1$ \\
$T_{\rm dust}$ [K] ($\beta = 2.0$) & $39.2 \pm 1.5$ & $ 34.8\pm 1.2$\\
$\log{\left(L_{\rm IR} / L_\odot\right)}$ & $12.7 \pm 0.1$ & $12.5 \pm 0.1$\\
${\rm SFR \, [M_\odot \, yr^{-1}]}$ & $\sim 900$ & $\sim 600$\\
\hline

\end{tabular}
\end{table}

\section{Dust morphology on 150 pc scales}\label{section_morphology_pc_scales}

\begin{table*}[!t]
\caption{\label{properties_SF_clumps} Observed properties of the star-forming clumps found in A--1 and A--2}
\centering
\begin{tabular}{cccccccc}
\hline
Clump & $S_{\rm 920 \, \mu m}$ & $\log{\left(L_{\rm IR} / L_\odot\right)}$\tablenotemark{a} & SFR & $A_{\rm d} \tablenotemark{b}$ & $A_{\rm d} \tablenotemark{b}$ & $\Sigma_{\rm IR}\tablenotemark{c}$ & $\Sigma_{\rm SFR}$ \\
 & [mJy] & & [${\rm M}_\odot \, {\rm yr}^{-1}$] & [${\rm mas} \times {\rm mas}$] & [${\rm pc} \times {\rm pc}$] & [$L_\odot \, {\rm kpc}^{-2}$]  & [${\rm M}_\odot \, {\rm yr}^{-1} \, {\rm kpc}^{-2}$]\\
\hline\hline
A--1A & $2.0 \pm 0.1$ 		&$12.2 \pm 0.2$ 	& $\sim 310$ 	& $49 \pm 5 \times 25 \pm 3$ 	& $360 \pm 40 \times 180 \pm 20$ & $\sim 1.7 \times 10^{13}$ & $\sim 3015$\\
A--1B & $0.9 \pm 0.1$		&$11.9 \pm 0.2$ 	& $\sim 140$ 	& $57 \pm 9 \times 25 \pm 9$ 	& $420 \pm 70 \times 180 \pm 70$ & $\sim 0.7 \times 10^{13}$ & $\sim 1180$\\
A--2A & $2.1 \pm 0.2$		&$12.3 \pm 0.2$ 	& $\sim 330$ 	& $57 \pm 5 \times 34 \pm 4$ 	& $420 \pm 40 \times 250 \pm 30$ & $\sim 1.2 \times 10^{13}$ & $\sim 2035$\\
\hline
\hline
\tablenotetext{1}{The total IR luminosities have been calculated from the observed flux density at $920 \, {\rm \mu m}$ assuming the best-fit FIR SEDs obtained from the mid-resolution data (see Figure \ref{figure_TDLIR_components}). Furthermore, we assume that all IR luminosity is due to star formation rather than AGN activity.}
\tablenotetext{2}{The sizes reported in the table correspond to the beam--deconvolved FWHM of a 2D elliptical Gaussian fit. We only report the size of the clumps detected at ${\rm SNR > 10}$.}
\tablenotetext{3}{The surface densities have been calculated assuming that the size of the sources is $\pi R_{\rm a} \times R_{\rm b}$, where $R_{\rm a}$ and $R_{\rm b}$ are the semi-axis of the best-fit elliptical Gaussian of each component.}
\end{tabular}
\end{table*}

\begin{figure*}
\centering
\includegraphics[width=0.50\textwidth]{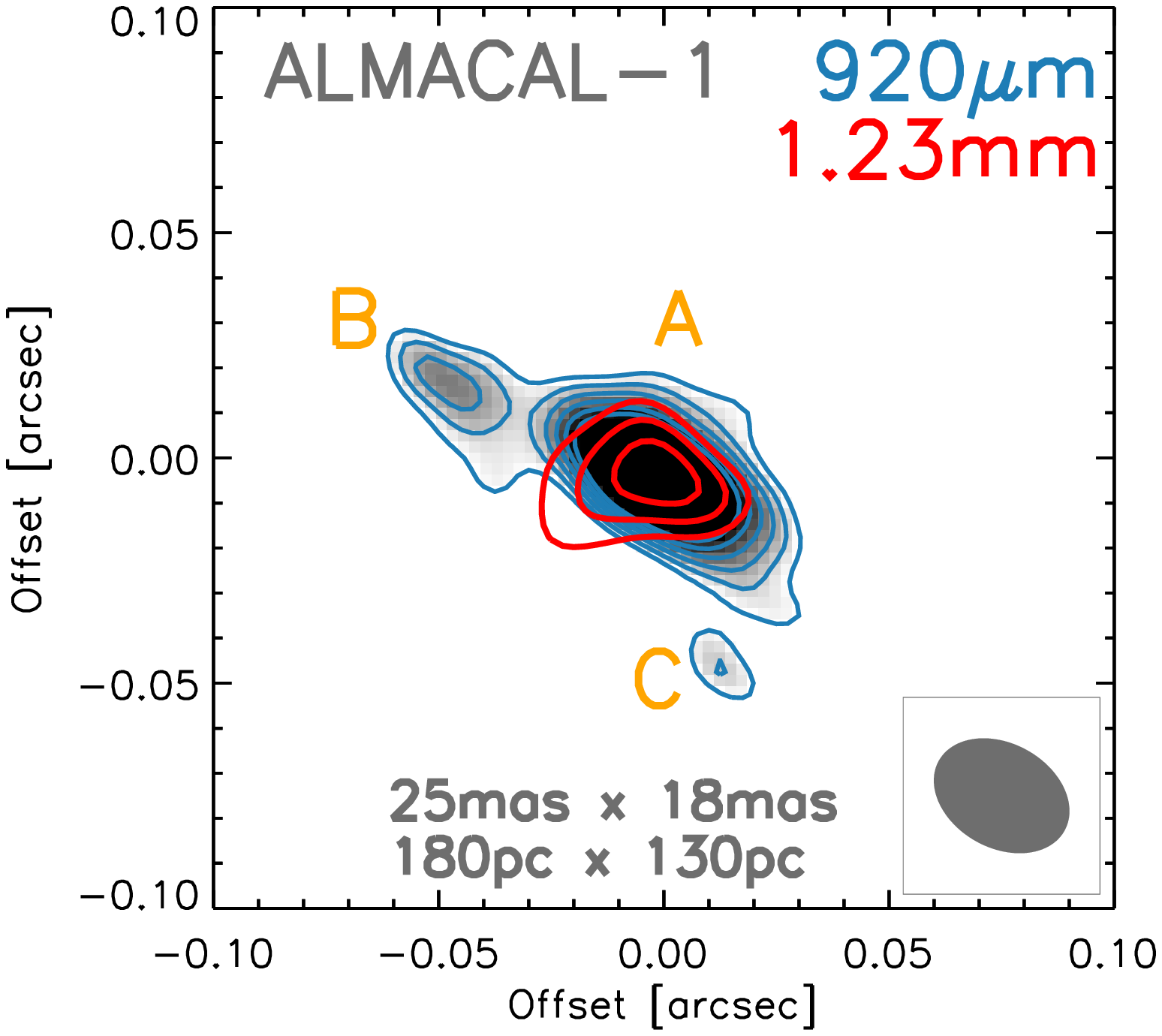}
\hspace{-5mm}
\includegraphics[width=0.50\textwidth]{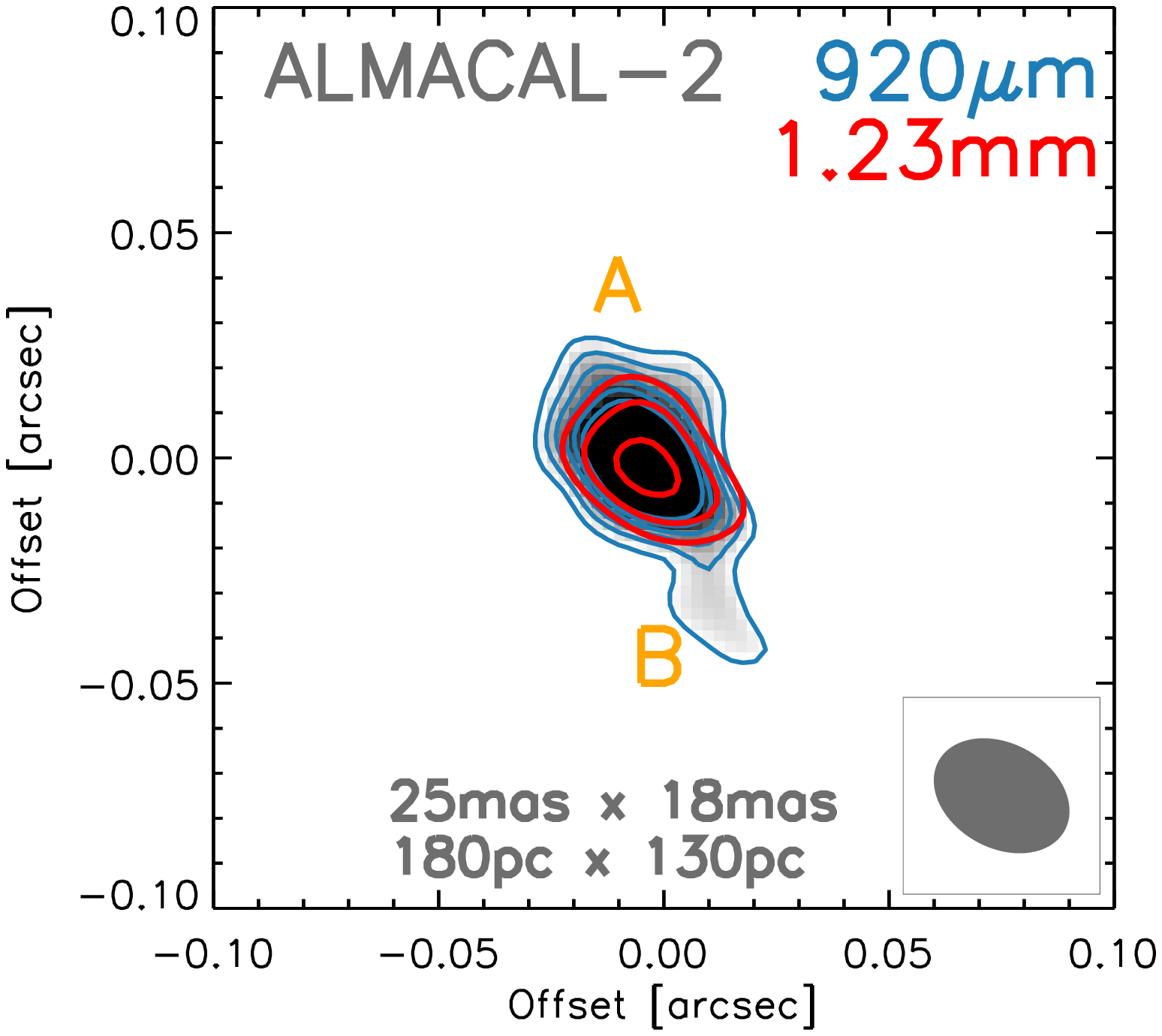}
\caption{Ultra-high-resolution imaging of ALMACAL--1 (A--1: {\it left}) and ALMACAL--2 (A--2: {\it right}). The background images and blue contours represent the ${\rm 920 \, \mu m}$ emission, while red contours are ${\rm 1.2 \, mm}$ emission. The synthesized beam and its size, both in sky and physical units, are indicated on each panel. All contours are represented from $5\sigma$, in steps of $1 \sigma$. It should be noted that the spatial resolution of our observations are about 10$\times$ times better than previous observations of high-redshift unlensed starbursts, and only comparable with the source-plane resolution of the ALMA long-baseline observations of SDP.81 \citep{Vlahakis2015ApJ...808L...4A} and the Eyelash \citep{Swinbank2010Natur.464..733S}. The dust emission in A--1 ({\it left}) is resolved into three different star-forming clumps (A--1A, A--1B, A--1C) while A--2 is resolved into two (A--2A and A--2B). The flux density of each clump in combination with their sizes reveal SFR surface densities significantly higher than those reported so far in high-redshift starbursts. Note that the size of each image is only $0.2''$ on each side.
              }
\vspace{8mm}
\label{hires_imaging_ALMA}
\end{figure*}

We focus in this section on the analysis of the dust continuum emission detected in B7 ($\sim 920 \, {\rm \mu m}$) and B6 ($\sim 1.23 \, {\rm mm}$) in our two dusty starbursts. A--1, the most luminous component of the pair (Figure \ref{hires_imaging_ALMA} -- left), is resolved into three star-forming clumps, A--1A, A--1B, and A--1C, with A--1A being more than 2$\times$ times brighter than the other two components. Only one star-forming clump (A--2A) is detected in A--2 with the r.m.s. of our data. We have measured the primary-beam corrected flux density at $920 \, {\rm \mu m}$ and beam--deconvolved size of each component with the task {\sc imfit} within {\sc casa}. The derived values are quoted in Table \ref{properties_SF_clumps}, where we only include the three clumps which are detected at $> 10\sigma$, since sizes cannot be reliably measured at lower signal to noise. We see that components A--1A and A--2A dominate the dust emission in A--1 and A--2, respectively, and appear very compact, with FWHM sizes of $\sim 300 \, {\rm pc}$. 

Figure \ref{hires_imaging_ALMA} also shows the ultra--high--spatial--resolution $1.2 \, {\rm mm}$ emission in our pair of SMGs. At $z = 3.442$, this wavelength probes the emission at rest-frame $\sim 280 \, {\rm \mu m}$. The spatial resolution of the B6 observations ($29 \, {\rm mas} \times 25 \,{\rm mas}$) is slightly worse than the resolution of the $920 \, {\rm \mu m}$ observations, but still comparable. The maximum of the $1.23 \, {\rm mm}$ emission (detected in components A--1A and A--2A only due to the sensitivity of the $1.23 \, {\rm mm}$ observations) is coincident with the maximum of the $920 \, {\rm \mu m}$ emission. However, there seems to be an elongation of the $1.23 \, {\rm mm}$ emission in A--1A which is not seen in the $920 \, {\rm \mu m}$ map. The origin of this extended emission is currently unknown, although further data on this source providing higher signal to noise detections or better $uv$ coverage will help to explore this issue further. In any case, the similarity between the B6 and B7 emission in A--1A and A--2A confirm that the dust emission in those two components is truly compact. Using the best-fit FIR SEDs of A--1 and A--2 (see Figure \ref{figure_TDLIR_components}) we estimate that their flux densities at $920 \, {\rm \mu m}$ are $S_{\rm 920 \, \mu m} = 5.3 \pm 0.2\, {\rm mJy}$ and $S_{\rm 920 \, \mu m} = 3.5 \pm 0.2 \, {\rm mJy}$, respectively. Considering the observed flux densities in our ultra-high-resolution data (see Table \ref{properties_SF_clumps}), we estimate that we resolve out about 40\% of the observed flux in both A--1 and A--2. This suggests that a significant fraction of the dust emission in our two dusty starbursts is relatively diffuse and/or extended, but that $\sim 60\%$ of the dust emission in A--1 and A--2 is extremely compact.


The total IR luminosity of each clump has been obtained by re-scaling the best-fit mid-IR power law plus the optically thin dust emission to A--1 and A--2 to their observed $920 \, {\rm \mu m}$ flux density. The uncertainties of the total IR luminosities are the same for all clumps and reflect the errors in the extrapolation from a single--band photometry to the total IR luminosity (changes in dust temperature, dust emissivity, power law of the mid-IR SED, etc). The associated SFRs have been derived using the classical \cite{Kennicutt1998} calibration and assuming a Salpeter IMF. The SFR of our star-forming clumps range from $80$ to $300 \, {\rm M}_\odot \, {\rm yr}^{-1}$. It is notable that the high SFR in A--1 and A--2 (especially in A--1A and A--2A) is taking place in extremely small star-forming clumps, with average FWHM sizes of about $300\, {\rm pc}$. This means that the SFR surface density, $\Sigma_{\rm SFR}$, of the clumps is as high as $\Sigma_{\rm SFR} \sim 3000 \, {\rm M}_\odot \, {\rm yr}^{-1} \, {\rm kpc}^{-2}$ (see Table \ref{properties_SF_clumps}). Such high values of the SFR surface densities have not been reported so far in any high-redshift dust starburst, and exceed the maximum value predicted by \cite{Andrews2011ApJ...727...97A}, $\sim 1000 \, {\rm M}_\odot \, {\rm yr}^{-1} \, {\rm kpc}^{-2}$.

\cite{Simpson2015ApJ...799...81S} reported a median value of $90 \pm 30 \, M_\odot \, {\rm yr}^{-1} \, {\rm kpc}^{-2}$ for their SMGs, with only two galaxies above $500 \, {\rm M}_\odot \, {\rm yr}^{-1} \, {\rm kpc}^{-2}$. In SGP38326, the most luminous, unlensed starburst found at $z > 4$, the star formation is taking place in two interacting disks, with the SFR rate density of the most luminous component of the merger being $\sim 840 \, M_\odot \, {\rm yr}^{-1} \, {\rm kpc}^{-2}$ \citep{Oteo2016arXiv160107549O}. For the Eyelash, a strongly lensed starburst at $z \sim 2.3$ whose star formation is occurring in four distinct clumps (${\rm FWHM \sim 100 - 300\,{\rm pc}}$), \cite{Thomson2015MNRAS.448.1874T} derived values as high as $\Sigma_{\rm SFR} \sim 1650 \, M_\odot \, {\rm yr}^{-1} \, {\rm kpc}^{-2}$. Other extreme dusty starbursts at high redshift have high SFR surface densities, such as HFLS3 ($\sim 600 \, M_\odot \, {\rm yr}^{-1} \, {\rm kpc}^{-2}$) or AzTEC-3 ($\sim 850 \, M_\odot \, {\rm yr}^{-1} \, {\rm kpc}^{-2}$) but none of them comparable to the values found in A--1 and A--2. Relatively low values of the SFR surface density are also found in extreme high-redshift galaxies, such as HDF\,850.1, with $\Sigma_{\rm SFR} \sim 35 \, M_\odot \, {\rm yr}^{-1} \, {\rm kpc}^{-2}$. It is clear that there is a significant variety of $\Sigma_{\rm SFR}$ values in high-redshift galaxies, or more likely that most data do not resolve the small star-forming clumps.

The reason for all previous $\Sigma_{\rm SFR}$ at high-redshift being significantly lower than in the star-forming clumps of A--1 and A--2 is likely a combination of their brightness and the availability of ultra--high--spatial--resolution observations revealing that the strong star formation is occurring in very small scales. Most previous work on unlensed SMGs employ observations with a linear resolution around 10$\times$ times worse than the resolution of our ALMA data. To highlight the importance of ultra--high--spatial--resolution observations in the analysis of the ISM of dusty starbursts we have determined the size of the dust emission and the derived value of $\Sigma_{\rm SFR}$ in A--1 and A--2 by using our mid-resolution observations (see Figure \ref{fig_map_calibrator} and \S \ref{data_set_ALMACAL}). The smaller beam is provided by the B9 observations, $0.50'' \times 0.30''$. With this, A--1 has a beam-deconvolved size of $345 \, {\rm mas} \times 194 \, {\rm mas}$, or $2.5 \, {\rm kpc} \times 1.4 \, {\rm kpc}$. This would imply $\Sigma_{\rm SFR} \sim 165 \, M_\odot \, {\rm yr}^{-1} \, {\rm kpc}^{-2}$, compatible with the values found by \cite{Simpson2015ApJ...799...81S} but more than one order of magnitude lower than the $\Sigma_{\rm SFR}$ of any of the three components A--1 is resolved into when observed at ultra--high--spatial resolution.

\cite{Wilson2014ApJ...789L..36W} reported high-resolution observations of the two nuclei of Arp 220 at $\sim 0.3'' \times 0.2''$. In physical scale, their spatial resolution ($\sim 130 \,{\rm pc} \times 70 \, {\rm pc}$) is matching our ultra--high--spatial resolution. \cite{Wilson2014ApJ...789L..36W} obtained IR surface density of $\Sigma_{\rm IR} = 2.1 \times 10^{14} \, {\rm L_\odot} \, {\rm kpc}^{-2}$ and $\Sigma_{\rm IR} = 5.8 \times 10^{12} \, {\rm L_\odot} \, {\rm kpc}^{-2}$ for the western and eastern nuclei, respectively. These values are similarly high to the values we find for A--1 and A--2 and were obtained at similar physical spatial resolution, highlighting again that ultra-high-spatial resolution plays a key role in the understanding the properties and nature of dusty starbursts.


It is possible that some of the observed $L_{\rm IR}$ in our two SMGs might be due to dust heated by an AGN in the center of the galaxies instead of star formation. In fact, \citep{Wilson2014ApJ...789L..36W} discussed that the extremely high luminosity surface densities found in the western nucleus of Arp 220 could be, in part, due to the presence of an AGN. If there is AGN contribution to the luminosity of the brightest clumps in A--1 and A--2, their SFRs would be overestimated, and so the associated SFR surface densities. It could be expected that, if there is AGN contribution in our two SMGs, the AGN is located in the two brightest clumps, but not in the fainter ones. Therefore, even if the SFR surface density of the brightest clumps might be overestimated, this is less likely to happen in the fainter clumps, and these still have high SFR surface densities. This assumes that the possible AGN heat the dust locally over a scale of less than $200 \, {\rm pc}$, not reaching the other star-forming clumps. 

\subsection{Exploring possible lensing}\label{lensing_section_discussion}

As commented in \S \ref{section_redshift_confirmation_J1058}, the fact that the flux ratio (both line and continuum) between A--1 and A--2 depends on wavelength indicates that these sources are not two gravitationally amplified components of a galaxy at $z = 3.442$ close to the line of sight of the blazar host. Despite this, and in order to investigate what the consequences of lensing would be, we consider in this section the possibility that A--1/A--2 is actually a lensed system. If this is the case, we need to calculate the flux and size of the source in the source plane. To do this we have used the code \verb+uvmcmcfit+, which models the lensed emission of galaxies observed with interferometers in the $uv$ plane \citep{Bussmann2013ApJ...779...25B,Bussmann2015ApJ...812...43B}. In \verb+uvmcmcfit+ the background source is assumed to have an elliptical Gaussian profile, whereas the lens mass profile is represented by a singular isothermal ellipsoid. 

We have first modeled the possible lensed emission in the mid-resolution maps (see Figure \ref{fig_map_calibrator}) with the aim to explore whether the spatial configuration of A--1 and A--2 with respect to the lens can be successfully modeled. We have modeled the lensed emission in all bands where A--1 and A--2 are detected (see Table \ref{table_properties_J1058_SMGs}). The result is that the positions and flux ratios of the two sources are well recovered in all bands. It should be noted that we have not modeled the multi-band emission simultaneously since this is not possible to do with publicly available codes working in the $uv$ plane. The magnification factor is derived from the ratio between the total flux density in the lensed image of the model to the total flux density in the unlensed, intrinsic source model. We have derived $\mu_{\rm dust} = 7.44 +/- 0.04$ at $870 \, {\rm \mu m}$, and similar values are obtained in the other bands (expected due to the similar spatial configuration of the system in the different bands). The effective radius of the source in the source plane is $R_{\rm eff} \sim 340 \, {\rm pc}$. We then used the best-fit model obtained from the mid-resolution data as an initial condition to model the possible lensed emission in the ultra-high-resolution observations. If they were lensed, the observed emission in A--1 and A--2 (see Figure \ref{hires_imaging_ALMA}) is compatible with a single, extremely compact background source, whose effective radius is only $R_{\rm eff} \sim 40 \, {\rm pc}$. 

The total observed SFR of A--1 and A--2 in the ultra-high-resolution observations is $ {\rm SFR} \sim 870 \, M_\odot \, {\rm yr^{-1}}$, which means a source-plane, de-magnified SFR of $\sim 120 \, {\rm M_\odot \, {\rm yr}^{-1}}$. Together with the effective radius in the source plane, the de-magnified SFR surface density is $\Sigma_{\rm SFR} \sim 10,000 M_\odot \, {\rm yr}^{-1} \, {\rm kpc}^{-2}$. This value is considerably higher than the values obtained considering that A--1 and A--2 are not lensed, and therefore, much higher than in any previous high-redshift source and very close to the value found in the eastern nucleus of Arp 220.


\section{Conclusions}\label{concluuuuu}

In this paper we have presented ultra-high spatial resolution ($\sim 20 \, {\rm mas}$) dust continuum ($870 \, {\rm \mu m}$ and $1.2 \, {\rm mm}$) observations of two dusty starbursts, the brightest SMGs detected so far in our survey of serendipitous sources in the fields of ALMA calibrators: A--1 ($S_{\rm 870 \mu m} = 6.5 \pm 0.2 \, {\rm mJy}$) and A--2 ($S_{\rm 870 \mu m} = 4.4 \pm 0.2 \, {\rm mJy}$). The main conclusions of our work are the following:

\begin{enumerate}

	\item We have determined the spectroscopic redshift of our two dusty starbursts to be $z = 3.442$ via detection of up to nine $^{12}$CO and ${\rm H_2O}$ emission lines in ALMA bands 4, 6, and 7. The maximum velocity shift found between the emission lines of A--1 and A--2 (which are separated on the sky by $28 \, {\rm kpc}$) is less than $100 \, {\rm km \, s^{-1}}$, significantly lower than in other high-redshift interacting starbursts.
		
	\item Using flux densities measured in ALMA band 6, 7, 8, and 9 we have determined the dust temperature and total IR luminosity of each of the two dusty starbursts. These values are compatible with those found for the classical population of SMGs (with A--1 being warmer than A--2), and they would have been selected as SMGs in single-dish submm surveys. Uniquely, the FIR SEDs of our two dusty starbursts have been constrained with sub-arcsec resolution observations, unlike in previous work based on single-dish FIR/submm observations, which suffer from large beam sizes and source confusion problems.
	
	\item Our ALMA ultra-high-resolution imaging reveals that about half of the dust emission in A--1 and A--2 is arising in compact components (with FWHM sizes of $\sim 350 \, {\rm pc}$). Two additional, fainter star-forming clumps are found in A--1. We recall that our in-field calibrator and subsequent self-calibration ensures near-perfect phase stability on the longest baselines, ensuring great image quality. Actually, we have two independent datasets in ALMA B6 and B7 at similar spatial resolution which prove the reliability of the reported structures.  
	
	\item The high SFR and the compact size of all the star-forming clumps in A--1 and A--2 indicate extremely high SFR surface densities of up to $\Sigma_{\rm SFR} \sim 6,000 \, M_\odot \, {\rm yr}^{-1} \, {\rm kpc}^{-2}$. These values are significantly higher than those previously obtained in high-redshift dusty starbursts, and only comparable to the values found in the nuclei of Arp 220 with observations at similar (physical) spatial resolution. It should be noted that the SFR is obtained assuming that the IR luminosity is due to star formation, since with the current data we cannot study what the contribution of a possible AGN to the SFR could be.
	
	\item We argue that the extremely high SFR surface densities of the star-forming clumps in A--1 and A--2 might be common in high redshift dusty starbursts but are only visible thanks to the availability of ultra-high spatial resolution data. This highlights the importance of long-baseline observations for the study of the ISM of dusty-starburst in the early Universe.
	
	\item There is a small probability that this system is lensed, in the sense that the two SMGs around J1058+0133 are actually the lensed emission of a source gravitationally amplified by the blazar host. If this is actually the case, the resolution of the observations would increase to $\sim 50 \, {\rm pc}$ and we would be resolving sizes comparable to individual giant molecular clouds. The galaxy in the source plane would have an effective radius of $R_{\rm eff} \sim 40 \, {\rm pc}$, implying a de-magnified SFR surface density of $\Sigma_{\rm SFR} \sim 10,000 \, M_\odot \, {\rm yr}^{-1} \, {\rm kpc}^{-2}$, only comparable with the value found in the eastern nucleus of Arp 220. 

\end{enumerate}

\begin{acknowledgements}
IO and RJI acknowledge support from the European Research Council in the form of the Advanced Investigator Programme, 321302, {\sc cosmicism}. We acknowledge stimulating discussions with Zhi-Yu Zhang. IRS acknowledges support from STFC (ST/L00075X/1), the ERC Advanced Investigator programme DUSTYGAL 321334 and a Royal Society/Wolfson Merit Award. This paper makes use of the following ALMA data: ADS/JAO.ALMA\#2015.1.01518.S, ADS/JAO.ALMA\#2015.1.00695.S, ADS/JAO.ALMA\#2015.1.00686.S, ADS/JAO.ALMA\#2015.1.01302.S, ADS/JAO.ALMA\#2015.1.01345.S, ADS/JAO.ALMA\#2015.1.00607.S, ADS/JAO.ALMA\#2015.1.00597.S, ADS/JAO.ALMA\#2015.1.00928.S, ADS/JAO.ALMA\#2015.1.00665.S. ALMA is a partnership of ESO (representing its member states), NSF (USA) and NINS (Japan), together with NRC (Canada) and NSC and ASIAA (Taiwan) and KASI (Republic of Korea), in cooperation with the Republic of Chile. The Joint ALMA Observatory is operated by ESO, AUI/NRAO and NAOJ.

\end{acknowledgements}

\bibliographystyle{mn2e}

\bibliography{ioteo_biblio}

\end{document}